\documentclass[runningheads]{llncs}
\usepackage[T1]{fontenc}
\usepackage{graphicx}
\usepackage{tcolorbox}
\usepackage{amsmath} 
\usepackage{booktabs}
\usepackage{multirow} 
\usepackage{array}

\usepackage{hyperref}
\usepackage{graphicx}
\usepackage{subcaption}
\usepackage{algorithm}
\usepackage{algpseudocode}

\begin{document}
\title{Cross-Domain Recommendation Meets Large Language Models}
%
%

\author{Ajay Krishna Vajjala \and
Dipak Meher \and
Ziwei Zhu \and David S. Rosenblum}
\authorrunning{Krishna Vajjala et al.}
%
\institute{George Mason University, Virginia, USA\\
\email{\{akrish,dmeher,zzhu20,dsr\}@gmu.edu}}
\maketitle              
\begin{abstract}
Cross-domain recommendation (CDR) has emerged as a pro\-mising solution to the cold-start problem, faced by single-domain recommender systems. However, existing CDR models rely on complex neural architectures, large datasets, and significant computational resources, making them less effective in data-scarce scenarios or when simplicity is crucial. In this work, we leverage the reasoning capabilities of large language models (LLMs) and explore their performance in the CDR domain across multiple domain pairs. We introduce two novel prompt designs tailored for CDR and demonstrate that LLMs, when prompted effectively, outperform state-of-the-art CDR baselines across various metrics and domain combinations in the rating prediction and ranking tasks. This work bridges the gap between LLMs and recommendation systems, showcasing their potential as effective cross-domain recommenders.

\end{abstract}

\section{Introduction}

Recommender systems (RS) have been widely adopted by e-commerce platforms to sift through large number of items and provide personalized recommendations specific to individual preferences \cite{zhao2023recommender}. However, most RS operate within a single domain, meaning they are trained on data from one specific domain and are limited to making recommendations within that domain \cite{vajjala2024vietoris}. These single-domain RS struggle with the cold-start problem, which arises when making recommendations for new users who lack interaction history \cite{vajjala2024analyzing}. 

To address the challenges faced by single-domain RS, cross-domain recommendation (CDR) has emerged. CDR leverages information from a “source domain” to make recommendations in a “target domain” \cite{zhu2022personalized,man2017cross}. When users have no or limited interactions in the target domain, CDR transfers their rating patterns from the source domain to create more accurate user representations in the target domain, leading to improved recommendations \cite{cao2023towards}. However, existing CDR approaches face several limitations. First, many models rely on complex neural architectures trained on relatively small datasets, which limits their reasoning ability and restricts them to memorizing explicit patterns without deeper inference capabilities \cite{hu2018conet,zhu2022personalized}. Second, while some models incorporate content-based features, the knowledge they gain about items is limited compared to the contextual understanding in existing pre-trained models \cite{vajjala2024vietoris,petruzzelli2024instructing}. Third, CDR models have no reasoning capabilities and rely solely on the data they are trained on, which makes them unable to generalize beyond the information provided in the training data. These constraints can impact CDR performance in scenarios where data is sparse or computational resources are constrained.

Recently, large language models (LLMs) have demonstrated impressive reasoning capabilities in machine learning research \cite{naveed2023comprehensive}. In addition, they have demonstrated impressive performance in domain adaptation tasks, often competing with state-of-the-art methods across various domains \cite{guo2022domain}. Their ability to generalize across tasks without task-specific training makes them a reasonable alternative for the cross-domain recommendation problem. In addition to their reasoning capabilities, LLMs offer significant advantages in handling textual and contextual information, which are often important pieces of side information for recommendation tasks \cite{zhao2023recommender}. Unlike traditional models that require domain-specific embeddings and features, LLMs can understand textual descriptions of items, user reviews, and metadata, which allows them to bridge gaps between domains using natural language \cite{petruzzelli2024instructing}. In situations where data is sparse, LLMs have the ability to infer preferences based on subtle patterns in user behavior. 

The motivation behind this work is to investigate whether LLMs, with their reasoning capabilities, can effectively handle CDR tasks and achieve performance on par with or better than existing models, even in scenarios with limited data and computational resources. In this work, we leverage LLMs for cross-domain recommendation and evaluate their performance in rating and ranking tasks compared to state-of-the-art methods. This work is important for CDR as it highlights the potential of LLMs to enhance recommendation performance by leveraging their ability to understand and transfer user preferences across domains. Finally, it contributes to the field of LLMs by demonstrating their versatility in addressing challenges in personalized recommendation, which further bridges the gap between natural language understanding and domain-specific tasks. We make our code and data available at \href{https://github.com/ajaykv1/CDR_Meets_LLMs}{https://github.com/ajaykv1/CDR\_Me\-ets\_LLMs}. To summarize, the contributions are as follows:

\begin{itemize}
    \item We introduce a prompting framework for LLMs specific to the CDR task, and provide insights into the key components of an effective prompt for CDR.
    \item We present two types of prompts for CDR and show their effectiveness: (i) incorporates user interactions from both the source and target domains; (ii) incorporates interactions exclusively from the source domain.
    \item We conduct an extensive evaluation of LLM performance on ranking and rating tasks across three domain combinations, and demonstrate that LLMs outperform baseline CDR models across various domain pairs.
\end{itemize}

\section{LLMs for Cross-Domain Recommendation}

We present two types of prompt strategies for the CDR domain: (i) \textbf{Target Domain Behavior Injection}; (ii) \textbf{No Target Domain Behavior Injection}. The "target domain behavior injection" prompt includes the user's interaction history from both the source and target domains, simulating a warm-start scenario. In contrast, the "no target domain behavior injection" prompt includes only the user's interaction history from the source domain, excluding target domain interactions, simulating a cold-start scenario. This difference in prompts not only helps evaluate the LLMs' adaptability in warm-start and cold-start scenarios but also provides insight into their ability to find patterns from only source domain interactions. Analyzing the performance under these conditions can help to provide a better understanding of the LLM's reliance on target domain data and their ability to generalize across domain combinations. This approach not only provides a deeper understanding of the LLM's capabilities, but also lays the foundation for designing more robust recommendation systems that can work well across different CDR domain pairs. In addition, the insights from this work can lead to the design of more advanced prompt engineering techniques, potentially guiding the development of hybrid approaches that combine the strengths of LLMs and traditional CDR methods.

\subsection{Designing Effective Prompts for CDR}
\label{sec:design_prompt_cdr}

Given that recommender systems involve two primary tasks—rating prediction and ranking—we designed distinct prompts tailored to each task. Through careful prompt engineering and experimentation, we identified three key components essential for effective LLM prompts in CDR:

\begin{enumerate}
    \item \textbf{Defining the LLM’s Role:}
    The prompt must clearly establish the LLM’s role, such as stating, \textit{"You are a cross-domain recommender."} This helps the LLM assume a specific role and approach the task with the appropriate reasoning framework. By defining the role, the prompt sets clear expectations for how the model should process the input and formulate its output.
    
    \item \textbf{Clear Separation of Roles and Information:}
    The prompt should clearly separate the role definition from the input data, outlining domains, listing items, and associating user ratings with items. This structured design reduces ambiguity and helps the LLM focus on relevant details.
    
    \item \textbf{Explicit Task Definition:}
    The prompt should clearly specify the task. A well-defined task helps the LLM utilize the provided information effectively and produce outputs that align with the ground truth. Specifying the format of the output ensures consistency and reduces the risk of incorrect outputs.

\end{enumerate}

\subsubsection{Target Domain Behavior Injection}
\label{sec:target_domain_behavior_injection_prompt}

\begin{figure*}[t]
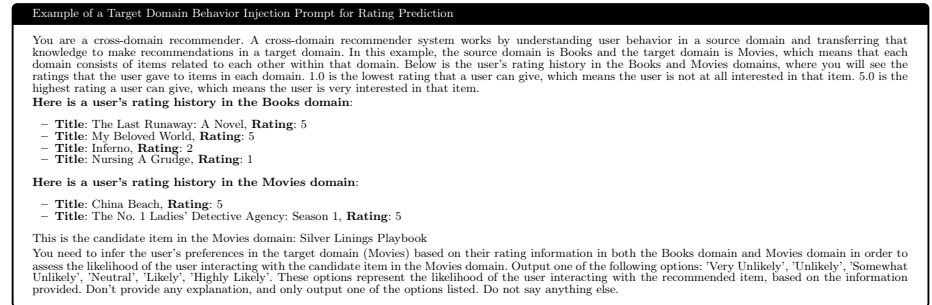
  
\centering
\scalebox{0.5}{
\begin{tcolorbox}[colback=white!10, colframe=black, width=2.0\textwidth, title=Example of a Target Domain Behavior Injection Prompt for Rating Prediction]
\renewcommand{\baselinestretch}{0.3}\selectfont  
\footnotesize  

You are a cross-domain recommender. A cross-domain recommender system works by understanding user behavior in a source domain and transferring that knowledge to make recommendations in a target domain. In this example, the source domain is Books and the target domain is Movies, which means that each domain consists of items related to each other within that domain. Below is the user’s rating history in the Books and Movies domains, where you will see the ratings that the user gave to items in each domain. 1.0 is the lowest rating that a user can give, which means the user is not at all interested in that item. 5.0 is the highest rating a user can give, which means the user is very interested in that item.\\

\textbf{Here is a user’s rating history in the Books domain}:
\begin{itemize}
    \item \textbf{Title}: The Last Runaway: A Novel, \textbf{Rating}: 5
    \item \textbf{Title}: My Beloved World, \textbf{Rating}: 5
    \item \textbf{Title}: Inferno, \textbf{Rating}: 2
    \item \textbf{Title}: Nursing A Grudge, \textbf{Rating}: 1
\end{itemize}
\textbf{Here is a user’s rating history in the Movies domain}:
\begin{itemize}
    \item \textbf{Title}: China Beach, \textbf{Rating}: 5
    \item \textbf{Title}: The No. 1 Ladies' Detective Agency: Season 1, \textbf{Rating}: 5
\end{itemize}

This is the candidate item in the Movies domain: Silver Linings Playbook\\

You need to infer the user’s preferences in the target domain (Movies) based on their rating information in both the Books domain and Movies domain in order to assess the likelihood of the user interacting with the candidate item in the Movies domain. Output one of the following options: 'Very Unlikely', 'Unlikely', 'Somewhat Unlikely', 'Neutral', 'Likely', 'Highly Likely'. These options represent the likelihood of the user interacting with the recommended item, based on the information provided. Don’t provide any explanation, and only output one of the options listed. Do not say anything else. 

\end{tcolorbox}
}
\vspace{-20pt}
\setlength{\belowcaptionskip}{-8pt} 
\caption{Target Domain Behavior Injection Prompt for Rating Prediction}
\label{box:prompt-with_injection_rating}  
\end{figure*}

\begin{figure*}[t]
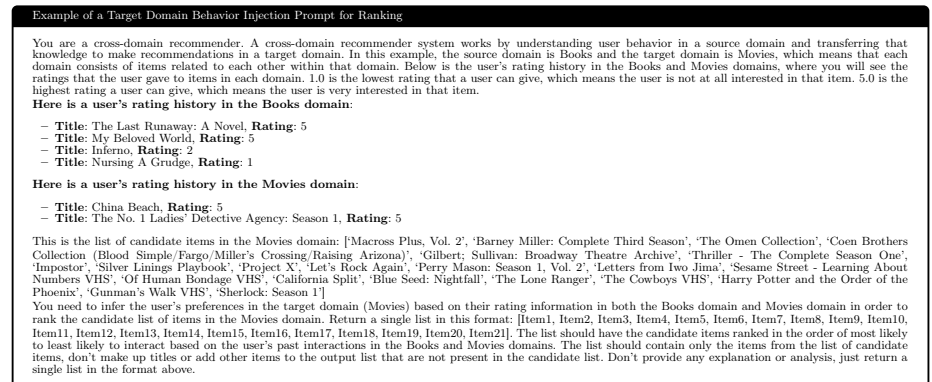
  
\centering
\scalebox{0.5}{
\begin{tcolorbox}[colback=white!10, colframe=black, width=2.0\textwidth, title=Example of a Target Domain Behavior Injection Prompt for Ranking]
\renewcommand{\baselinestretch}{0.3}\selectfont  
\footnotesize  

You are a cross-domain recommender. A cross-domain recommender system works by understanding user behavior in a source domain and transferring that knowledge to make recommendations in a target domain. In this example, the source domain is Books and the target domain is Movies, which means that each domain consists of items related to each other within that domain. Below is the user’s rating history in the Books and Movies domains, where you will see the ratings that the user gave to items in each domain. 1.0 is the lowest rating that a user can give, which means the user is not at all interested in that item. 5.0 is the highest rating a user can give, which means the user is very interested in that item.\\

\textbf{Here is a user’s rating history in the Books domain}:
\begin{itemize}
    \item \textbf{Title}: The Last Runaway: A Novel, \textbf{Rating}: 5
    \item \textbf{Title}: My Beloved World, \textbf{Rating}: 5
    \item \textbf{Title}: Inferno, \textbf{Rating}: 2
    \item \textbf{Title}: Nursing A Grudge, \textbf{Rating}: 1
\end{itemize}
\textbf{Here is a user’s rating history in the Movies domain}:
\begin{itemize}
    \item \textbf{Title}: China Beach, \textbf{Rating}: 5
    \item \textbf{Title}: The No. 1 Ladies' Detective Agency: Season 1, \textbf{Rating}: 5
\end{itemize}

This is the list of candidate items in the Movies domain: [`Macross Plus, Vol. 2', `Barney Miller: Complete Third Season', `The Omen Collection', `Coen Brothers Collection (Blood Simple/Fargo/Miller's Crossing/Raising Arizona)', `Gilbert; Sullivan: Broadway Theatre Archive', `Thriller - The Complete Season One', `Impostor', `Silver Linings Playbook', `Project X', `Let's Rock Again', `Perry Mason: Season 1, Vol. 2', `Letters from Iwo Jima', `Sesame Street - Learning About Numbers VHS', `Of Human Bondage VHS', `California Split', `Blue Seed: Nightfall', `The Lone Ranger', `The Cowboys VHS', `Harry Potter and the Order of the Phoenix', `Gunman's Walk VHS', `Sherlock: Season 1'] \\

You need to infer the user’s preferences in the target domain (Movies) based on their rating information in both the Books domain and Movies domain in order to rank the candidate list of items in the Movies domain. Return a single list in this format: [Item1, Item2, Item3, Item4, Item5, Item6, Item7, Item8, Item9, Item10, Item11, Item12, Item13, Item14, Item15, Item16, Item17, Item18, Item19, Item20, Item21]. The list should have the candidate items ranked in the order of most likely to least likely to interact based on the user’s past interactions in the Books and Movies domains. The list should contain only the items from the list of candidate items, don’t make up titles or add other items to the output list that are not present in the candidate list. Don't provide any explanation or analysis, just return a single list in the format above.

\end{tcolorbox}
}
\vspace{-20pt}
\setlength{\belowcaptionskip}{-18pt} 
\caption{Target Domain Behavior Injection Prompt for Ranking}
\label{box:prompt-with_injection_ranking}  
\end{figure*}

Traditional CDR approaches are typically trained using both source and target domain data \cite{zhu2022personalized,man2017cross,hu2018conet,zhu2021cross}. In this setup, models learn user interaction patterns across domains by training on thousands of examples, enabling them to capture user behavior and preferences. The trained model is then used to generate recommendations for users. To emulate this process, we designed a prompt that provides the LLM with a user’s interaction history in both the source and target domains (following the prompt structure from Section \ref{sec:design_prompt_cdr}). By incorporating both sources of information, the prompt encourages the LLM to identify patterns and recommend items in the target domain. Figures \ref{box:prompt-with_injection_rating} and \ref{box:prompt-with_injection_ranking} illustrate examples of this type of prompt, using Books as the source domain and Movies as the target domain for the rating and ranking tasks, respectively. In these prompts, we can observe the items rated by the user in the source domain, along with their explicit ratings, as well as the items rated in the target domain with explicit ratings. Additionally, the rating prediction prompt includes a candidate item to rate, while the ranking prediction prompt presents a candidate list of items to rank. We provide up to 10 historical interactions per user for both the source and target domains. This ensures compatibility with the context window of smaller LLMs, reduces potential confusion, and provides insight into LLM performance with limited examples.


\subsubsection{No Target Domain Behavior Injection}

\begin{figure*}[t]
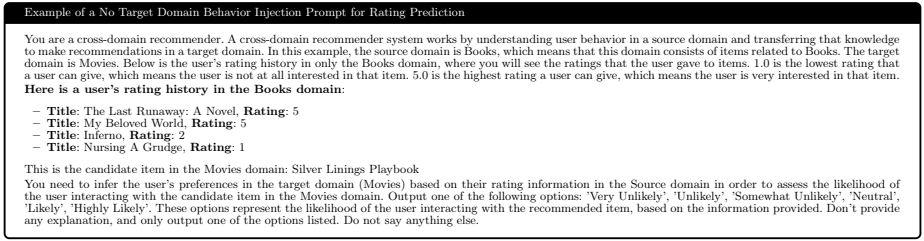
  
\centering
\scalebox{0.5}{
\begin{tcolorbox}[colback=white!10, colframe=black, width=2.0\textwidth, title=Example of a No Target Domain Behavior Injection Prompt for Rating Prediction]
\renewcommand{\baselinestretch}{0.3}\selectfont  
\footnotesize  

You are a cross-domain recommender. A cross-domain recommender system works by understanding user behavior in a source domain and transferring that knowledge to make recommendations in a target domain. In this example, the source domain is Books, which means that this domain consists of items related to Books. The target domain is Movies. Below is the user’s rating history in only the Books domain, where you will see the ratings that the user gave to items. 1.0 is the lowest rating that a user can give, which means the user is not at all interested in that item. 5.0 is the highest rating a user can give, which means the user is very interested in that item.\\

\textbf{Here is a user’s rating history in the Books domain}:
\begin{itemize}
    \item \textbf{Title}: The Last Runaway: A Novel, \textbf{Rating}: 5
    \item \textbf{Title}: My Beloved World, \textbf{Rating}: 5
    \item \textbf{Title}: Inferno, \textbf{Rating}: 2
    \item \textbf{Title}: Nursing A Grudge, \textbf{Rating}: 1
\end{itemize}

This is the candidate item in the Movies domain: Silver Linings Playbook\\

You need to infer the user’s preferences in the target domain (Movies) based on their rating information in the Source domain in order to assess the likelihood of the user interacting with the candidate item in the Movies domain. Output one of the following options: 'Very Unlikely', 'Unlikely', 'Somewhat Unlikely', 'Neutral', 'Likely', 'Highly Likely'. These options represent the likelihood of the user interacting with the recommended item, based on the information provided. Don’t provide any explanation, and only output one of the options listed. Do not say anything else.

\end{tcolorbox}
}
\vspace{-20pt}
\setlength{\belowcaptionskip}{-8pt} 
\caption{No Target Domain Behavior Injection Prompt for Rating Prediction}
\label{box:prompt-no_injection_rating}  
\end{figure*}

\begin{figure*}[t]
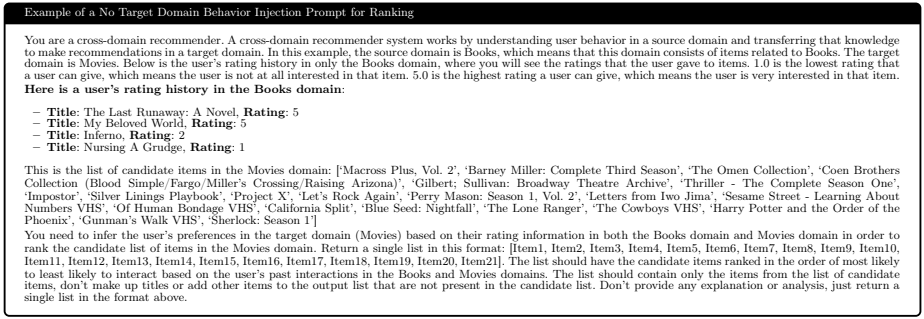
  
\centering
\scalebox{0.5}{
\begin{tcolorbox}[colback=white!10, colframe=black, width=2.0\textwidth, title=Example of a No Target Domain Behavior Injection Prompt for Ranking]
\renewcommand{\baselinestretch}{0.3}\selectfont  
\footnotesize  

You are a cross-domain recommender. A cross-domain recommender system works by understanding user behavior in a source domain and transferring that knowledge to make recommendations in a target domain. In this example, the source domain is Books, which means that this domain consists of items related to Books. The target domain is Movies. Below is the user’s rating history in only the Books domain, where you will see the ratings that the user gave to items. 1.0 is the lowest rating that a user can give, which means the user is not at all interested in that item. 5.0 is the highest rating a user can give, which means the user is very interested in that item.\\

\textbf{Here is a user’s rating history in the Books domain}:
\begin{itemize}
    \item \textbf{Title}: The Last Runaway: A Novel, \textbf{Rating}: 5
    \item \textbf{Title}: My Beloved World, \textbf{Rating}: 5
    \item \textbf{Title}: Inferno, \textbf{Rating}: 2
    \item \textbf{Title}: Nursing A Grudge, \textbf{Rating}: 1
\end{itemize}

This is the list of candidate items in the Movies domain: [`Macross Plus, Vol. 2', `Barney Miller: Complete Third Season', `The Omen Collection', `Coen Brothers Collection (Blood Simple/Fargo/Miller's Crossing/Raising Arizona)', `Gilbert; Sullivan: Broadway Theatre Archive', `Thriller - The Complete Season One', `Impostor', `Silver Linings Playbook', `Project X', `Let's Rock Again', `Perry Mason: Season 1, Vol. 2', `Letters from Iwo Jima', `Sesame Street - Learning About Numbers VHS', `Of Human Bondage VHS', `California Split', `Blue Seed: Nightfall', `The Lone Ranger', `The Cowboys VHS', `Harry Potter and the Order of the Phoenix', `Gunman's Walk VHS', `Sherlock: Season 1'] \\

You need to infer the user’s preferences in the target domain (Movies) based on their rating information in both the Books domain and Movies domain in order to rank the candidate list of items in the Movies domain. Return a single list in this format: [Item1, Item2, Item3, Item4, Item5, Item6, Item7, Item8, Item9, Item10, Item11, Item12, Item13, Item14, Item15, Item16, Item17, Item18, Item19, Item20, Item21]. The list should have the candidate items ranked in the order of most likely to least likely to interact based on the user’s past interactions in the Books and Movies domains. The list should contain only the items from the list of candidate items, don’t make up titles or add other items to the output list that are not present in the candidate list. Don't provide any explanation or analysis, just return a single list in the format above.

\end{tcolorbox}
}
\vspace{-20pt}
\setlength{\belowcaptionskip}{-15pt} 
\caption{No Target Domain Behavior Injection Prompt for Ranking}
\label{box:prompt-no_injection_ranking}  
\end{figure*}

To evaluate how LLMs perform with varying levels of information, we introduced a prompt that provides only the user's interaction history in the source domain, excluding interactions from the target domain. The task is to predict a rating or rank a list in the target domain using only source domain data, assessing the LLM's capability in such scenarios. Following the prompt structure outlined in Section \ref{sec:design_prompt_cdr}, we designed two prompts: one for rating prediction and one for ranking. Figures \ref{box:prompt-no_injection_rating} and \ref{box:prompt-no_injection_ranking} illustrate these prompts. The structure and information align with the "target domain behavior injection" prompt from Section \ref{sec:target_domain_behavior_injection_prompt}, but they use only source domain data. These prompts are designed to examine how LLMs perform with limited information when making recommendations in the target domain, simulating cold-start scenarios commonly encountered in traditional CDR methods. Similar to the target domain behavior injection prompt, we show a maximum of 10 historical interactions for the user in the source domain. 

\section{Experiments}

We conducted extensive experiments to evaluate LLMs for CDR, and aim to answer the following research questions: \textbf{RQ1}: How do LLMs perform in the ranking task for CDR compared to the state-of-the-art? \textbf{RQ2}: How do LLMs perform in the rating prediction task for CDR compared to the state-of-the-art? \textbf{RQ3}: How does the amount of context in a prompt affect results? Our code is available at \href{https://github.com/ajaykv1/CDR_Meets_LLMs}{https://github.com/ajaykv1/CDR\_Meets\_LLMs}.

\subsection{Dataset}

For experimentation, we adopted the Amazon Reviews dataset \cite{ni2019justifying}, which is widely used in the field of recommender systems (see Table \ref{tab:dataset_table}) \cite{zhu2022personalized,vajjala2024vietoris}. We chose five categories from this dataset to represent the domains used in this study. The domains we chose are Books, Electronics, Movies\_and\_Tv (Movies), Cds\_and\_Vinyl (Music), and Grocery\_and\_Gourmet\_Food (Food). From these domains, we create 3 CDR pairs: \textbf{Pair 1}: Books$\rightarrow$Movies, \textbf{Pair 2}:Movies$\-\rightarrow$Music, \textbf{Pair 3}: Electronics$\rightarrow$Food. The domain pairs were chosen to assess LLM performance across both closely related domains, such as Pairs 1 and 2, and completely unrelated domains, such as Pair 3. For each domain pair, we filtered the data to include only interactions from overlapping users present in both domains, and evaluated on 1000 interactions in the test dataset due to cost constraints.

\begin{table*}[t]
\centering
\renewcommand{\arraystretch}{0.9}
\scalebox{0.8}{
\begin{tabular}{c||c|c||c|c||c|c|c||c|c}
\toprule
\multirow{2}{*}{\textbf{CDR Pairs}} & \multicolumn{2}{c||}{\textbf{Domain Combination}} & \multicolumn{2}{c||}{\textbf{Item \#}} & \multicolumn{3}{c||}{\textbf{User \#}} & \multicolumn{2}{c}{\textbf{Rating \#}} \\
\cmidrule{2-3}\cmidrule{4-5}\cmidrule{6-8}\cmidrule{9-10}
& \textbf{Source} & \textbf{Target} & \textbf{Source} & \textbf{Target} & \textbf{Source} & \textbf{Target} & \textbf{Overlap} & \textbf{Source} & \textbf{Target} \\
\midrule
\textbf{Pair 1} & Books & Movies & 115172 & 50052 & 135109 & 123960 & 36705 & 828929 & 314123 \\
\textbf{Pair 2} & Movies & Music & 50052 & 64443 & 123960 & 75258 & 17534 & 314123 & 337221 \\
\textbf{Pair 3} & Electronics & Food & 63001 & 8713 & 192403 & 14681 & 5551 & 83060 & 39245 \\
\bottomrule
\end{tabular}
}
\setlength{\belowcaptionskip}{-15pt} 
\caption{Statistics from the Amazon Review Dataset, along with CDR tasks.}
\label{tab:dataset_table}
\end{table*}

\subsection{LLMs for CDR and Baseline Models}

In this work, we use both open-source Llama models \cite{touvron2023llama} and GPT models \cite{kalyan2023survey} to thoroughly evaluate LLM performance in CDR tasks. For the Llama models, we use llama-2-7b-chat, llama-2-13b-chat, and llama-3-8b-instruct. For the GPT models, we employ GPT-3.5-turbo, GPT-4, and GPT-4o, leveraging the latest and most advanced LLMs available \cite{islam2024gpt}. To evaluate LLM performance compared to the state-of-the-art CDR models, we chose the following baselines for both the rating and the ranking task. For the rating task, the baselines chosen are: 
\begin{itemize}
    \item \textbf{TGT} \cite{koren2009matrix}: A single-domain Matrix Factorization (MF) model trained exclusively on data from the target domain.
    \item \textbf{CMF} \cite{pan2010transfer}: A CDR model that extends MF by jointly factorizing rating matrices across multiple domains, using a shared global user embedding.
    \item \textbf{EMCDR} \cite{man2017cross}: A mapping-based CDR method that employs a multi-layer perceptron (MLP) as the general function to bridge domains.
    \item \textbf{PTUPCDR} \cite{zhu2022personalized}: A personalized bridge-based CDR model that integrates a meta-network with a linear mapping function to create bridges for users.
\end{itemize}

For the ranking taks, we chose the following baselines:
\begin{itemize}
    \item \textbf{PTUPCDR} \cite{zhu2022personalized}: We chose this model again for ranking, given that it was the best-performing model out of the baselines chosen for the rating task.
    \item \textbf{UniCDR} \cite{cao2023towards}: A unified CDR framework that leverages shared and domain-specific information to build a holistic representation of user preferences across domains, enabling more accurate predictions.
    \item \textbf{DisenCDR} \cite{cao2022disencdr}: A disentangled representation-based model for CDR, which learns independent latent factors for each domain while preserving shared user preferences, improving interpretability and performance.
\end{itemize}

\subsection{Evaluation Metrics}
\label{sec:evaluation_metrics}

We observed that LLMs struggle in predicting numerical ratings for the rating prediction task \cite{requeima2024llm,levy2024language}. To address this, we assigned textual labels to represent the range of possible ratings, from 0.5 to 5.0 in half-point increments: \textit{Very Unlikely}, \textit{Unlikely}, \textit{Somewhat Unlikely}, \textit{Neutral}, \textit{Likely}, and \textit{Highly Likely}. The LLM is instructed to output one of these labels, which we then map back to a numerical value in the rating prediction task. For the rating task, we chose the Root Mean Squared Error (RMSE) and the Mean Absolute Error (MAE) metrics \cite{hodson2022root} (shown in Equation \ref{eq:rating_metrics}). 

\setlength{\abovedisplayskip}{2pt}  
\setlength{\belowdisplayskip}{2pt}
{\scriptsize 
\begin{align}
    \text{RMSE} &= \sqrt{\frac{1}{N}\sum_{i=1}^{N}(y_i - \hat{y}_i)^2} \hspace{1cm} 
    \text{MAE} = \frac{1}{N}\sum_{i=1}^{N}|y_i - \hat{y}_i|
\label{eq:rating_metrics}
\end{align}
}

For the ranking task, we used the Mean Reciprocal Rank (MRR) and the Normalized Discounted Cumulative Gain (NDCG) metrics \cite{tamm2021quality} and evaluate them at the top 10 positions in the ranked list. 
{\scriptsize 
\begin{align}
    \text{NDCG} &= \frac{1}{|U_{\text{test}}|} \sum_{u \in U_{\text{test}}} \frac{\log 2}{\log(p_u + 1)} \hspace{1cm} 
    \text{MRR} = \frac{1}{|U_{\text{test}}|} \sum_{u \in U_{\text{test}}} \frac{1}{p_u}
\label{eq:ranking_metrics}
\end{align}
}

To perform the ranking task, we selected 20 negative samples for each positive item and calculated the metrics (as shown in Equation \ref{eq:ranking_metrics}) based on the position of the positive item in the ranked list \cite{ma2024negative}. 

\section{Results}

In this section, we present the results for the research questions and analyze how LLMs perform in CDR. For each LLM model, we use the labels "with" and "no" to indicate whether the prompt included "target domain behavior injection" or excluded it, respectively. For instance, "GPT-3.5-with" refers to results obtained using a prompt with target domain injection, while "GPT-3.5-no" corresponds to results obtained without target domain injection (see Tables \ref{tab:llm_ranking_results} and \ref{tab:llm_rating_results}). 

\subsection{RQ1: LLM Ranking Results for CDR}
\label{sec:ranking_results_section}

Table \ref{tab:llm_ranking_results} presents the performance of off-the-shelf LLMs in the ranking task for CDR. Due to the complexity of the prompts and the large context windows required, we utilized models from OpenAI's GPT family for this task. The results show that GPT-3.5, the smallest model, performs worse compared to its larger counterparts, GPT-4 and GPT-4o, which is expected given that GPT-4 and GPT-4o represent the state-of-the-art in performance \cite{kalyan2023survey}. From the results, we can see that the "with target domain behavior injection" prompts consistently outperform the "no target domain behavior injection" prompts. Among the LLMs used, GPT-4 achieves the best results for Pair 1, while GPT-4o performs the best in both Pair 2 and Pair 3. 

This is an encouraging result, as GPT-4 and GPT-4o, known as state-of-the-art models \cite{kalyan2023survey}, effectively leverage user interactions from both source and target domains to produce accurate ranking predictions in the target domain. Among the state-of-the-art baseline models for the ranking task in CDR, PTUPCDR performs best for Pair 1 and Pair 2, while UniCDR achieves the highest NDCG and DisenCDR the highest MRR for Pair 3. When comparing LLM performance with these baselines, GPT-4 surpasses the best baseline in Pair 1 with an MRR of 0.308 and NDCG of 0.383, compared to the baseline's 0.2596 and 0.3646. For Pair 2, GPT-4o outperforms the baseline with an MRR of 0.346 and NDCG of 0.412, exceeding the baseline's 0.2611 and 0.3822. This highlights the LLM's ability to generalize effectively from limited examples in intuitively similar domain combinations. For Pair 3, GPT-4o achieves the best MRR of 0.0219 and the best NDCG of 0.0244. These low values are expected since Pair 3 involves a domain combination that is not intuitively similar, using Electronics as the source domain and attempting to make recommendations in the Food domain. As anticipated, baseline CDR models outperform LLMs in Pair 3 across nearly all metrics. We limit the LLM prompt to contain very few source and target domain interactions (approximately 10), making it challenging for LLMs to generalize rating behavior across such dissimilar domains. 

These results demonstrate that for intuitively similar domain combinations (Pair 1 and Pair 2), LLMs can effectively leverage the provided rating history from the source and target domains to generalize well and outperform state-of-the-art baseline models in the CDR task. This finding is significant because it highlights that, despite the complexity of CDR, a detailed prompt can enable LLMs to achieve performance levels comparable to, or even surpassing, baseline models, making them a great alternative for CDR.

{\scriptsize
\begin{table*}[t]
\centering
\renewcommand{\arraystretch}{0.4}
\scalebox{0.85}{
\begin{tabular}{p{2.3cm}||>{\centering\arraybackslash}p{1.7cm}|>{\centering\arraybackslash}p{1.84cm}|>{\centering\arraybackslash}p{1.7cm}|>{\centering\arraybackslash}p{1.85cm}|>{\centering\arraybackslash}p{1.7cm}|>{\centering\arraybackslash}p{1.85cm}}
\toprule
\multirow{2}{*}{\textbf{Ranking Task}} & \multicolumn{2}{c|}{\textbf{Pair 1}} & \multicolumn{2}{c|}{\textbf{Pair 2}} & \multicolumn{2}{c}{\textbf{Pair 3}} \\
\cmidrule{2-7}
 & \textbf{MRR@10} & \textbf{NDCG@10} & \textbf{MRR@10} & \textbf{NDCG@10} & \textbf{MRR@10} & \textbf{NDCG@10} \\
\midrule
gpt-3.5-with & 0.303 & 0.379 & 0.342 & 0.408 & 0.0188 & 0.0219 \\
gpt-3.5-no & 0.233 & 0.316 & 0.221 & 0.296 & 0.0099 & 0.0123 \\
gpt-4-with & \textbf{0.308*} & \textbf{0.383*} & \textbf{0.341*} & 0.411 & 0.0191 & 0.0223\textbf{*} \\
gpt-4-no & 0.239 & 0.317 & 0.206 & 0.281 & 0.0108 & 0.0136 \\
gpt-4o-with & 0.305 & 0.378 & 0.339 & \textbf{0.412*} & 0.0193\textbf{*} & 0.0213 \\
gpt-4o-no & 0.239 & 0.318 & 0.229 & 0.294 & 0.0098 & 0.0118 \\
\midrule
\multicolumn{7}{c}{\textbf{Baseline Ranking Results}} \\
\midrule
PTUPCDR & 0.2596 & 0.3646 & 0.2611 & 0.3822 & 0.0902 & 0.1646 \\
UniCDR & 0.2171 & 0.2787 & 0.2127 & 0.2752 & 0.0184 & \textbf{0.2507} \\
DisenCDR & 0.1652 & 0.2866 & 0.1686 & 0.3085 & \textbf{0.1292} & 0.2008 \\
\bottomrule
\end{tabular}
}
\setlength{\belowcaptionskip}{-15pt} 
\caption{LLM and baseline results for ranking task across CDR pairs. \textbf{Bold} indicates the best performance and \textbf{*} indicates best LLM performance.}
\label{tab:llm_ranking_results}
\end{table*}
}

\subsection{RQ2: LLM Rating Results for CDR}

Table \ref{tab:llm_rating_results} presents the performance of off-the-shelf LLMs in the rating prediction task. For this task, we evaluate three models from the Llama family \cite{touvron2023llama}: the 7B and 13B chat models from Llama-2, and the 8B instruct model from Llama-3. From the GPT family, we use the same models as in the ranking task.

For the Llama models, the results indicate that the llama-3-8b-instruct model outperforms the other Llama models across all metrics and domain pairs. The llama-2-7b-chat model performs the worst, as expected, given that it is the smallest model with the fewest parameters \cite{galileo2023llama2}. Sources have also shown that the llama-3-8b-instruct model outperforms the llama-2-7b-chat and llama-2-13b-chat models across various NLP evaluation metrics \cite{neuralmagic2024llama3}, making its performance expected. Interestingly, the results reveal that for the llama-3-8b-instruct model, the "no target domain behavior injection" prompt performs better than the "with target domain behavior injection" prompt. This is an interesting finding, as the model achieves better performance when relying solely on source domain interactions without providing target domain interactions. 

We observe a similar pattern in the GPT models. GPT-3.5 performs the worst among all the GPT models, as expected, given that it is the smallest model in terms of parameters. The best-performing model is GPT-4o, which not only outperforms all other GPT models but also surpasses the Llama models. This result aligns with expectations, as GPT-4o is widely recognized as one of the best models and state-of-the-art in performance \cite{kalyan2023survey}. Interestingly, the same trend observed in the Llama models is also evident here: the best models perform better when using only the source domain data, without injecting target domain data. This finding highlights an intriguing behavior of LLMs in the rating prediction task for CDR, showcasing their ability to excel in this task using minimal domain-specific information. When compared to the state-of-the-art baselines for rating prediction, GPT-4o outperforms the baselines across all metrics and domain pairs, with the exception of MAE in Pair 1 and Pair 2. One surprising finding is that GPT-4o outperforms the baseline models in Pair 3. This result is particularly interesting, as the dissimilar nature of the domain combination in Pair 3 makes it challenging. However, the LLM identified generalizable patterns in the rating prediction task—unlike the ranking task—beating baseline models.

These results demonstrate that LLMs can outperform key baselines in the field of CDR for the rating prediction task in most cases, showing significant promise in generating relevant recommendations. An interesting finding is that utilizing only source domain behavior yields better results than incorporating target domain interaction behavior, which leads to suboptimal performance. This is a critical insight and presents an opportunity for researchers in the recommender systems and machine learning communities to explore further. Understanding the reasons behind this performance disparity and identifying ways to improve it is an important direction for future work. We emphasize this finding as a key contribution of our study and propose it as an open challenge for researchers to address in this domain.

{\scriptsize
\begin{table*}[t]
\centering
\renewcommand{\arraystretch}{0.4}
\scalebox{0.85}{
\begin{tabular}{p{2.5cm}||>{\centering\arraybackslash}p{1.2cm}|>{\centering\arraybackslash}p{1.2cm}|>{\centering\arraybackslash}p{1.2cm}|>{\centering\arraybackslash}p{1.2cm}|>{\centering\arraybackslash}p{1.2cm}|>{\centering\arraybackslash}p{1.2cm}}
\toprule
\multirow{2}{*}{\textbf{Model}} & \multicolumn{2}{c|}{\textbf{Pair 1}} & \multicolumn{2}{c|}{\textbf{Pair 2}} &
\multicolumn{2}{c}{\textbf{Pair 3}} \\
\cmidrule{2-7}
 & \textbf{MAE} & \textbf{RMSE} & \textbf{MAE} & \textbf{RMSE} & \textbf{MAE} & \textbf{RMSE} \\
\midrule
llama-2-7b-with & 1.849 & 2.377 & 1.872 & 2.361 & 1.946 & 2.371 \\
llama-2-7b-no & 1.862 & 2.396 & 1.904 & 2.394 & 1.959 & 2.385 \\
llama-2-13b-with & 1.589 & 1.973 & 1.624 & 2.019 & 1.646 & 1.981 \\
llama-2-13b-no & 1.682 & 2.101 & 1.743 & 2.208 & 1.715 & 2.161 \\
llama-3-8b-with & 1.611 & 2.037 & 1.704 & 2.137 & 1.583 & 1.941 \\
llama-3-8b-no & 1.491 & 1.859 & 1.367 & 1.729 & 1.489 & 1.817 \\
\midrule
gpt-3.5-with & 1.611 & 1.446 & 1.772 & 1.504 & 1.705 & 1.456 \\
gpt-3.5-no & 1.512 & 1.382 & 1.391 & 1.314 & 1.164 & 1.196 \\
gpt-4-with & 1.658 & 1.448 & 1.822 & 1.517 & 1.666 & 1.449 \\
gpt-4-no & 1.611 & 1.411 & 1.514 & 1.313 & 1.148 & \textbf{1.186*} \\
gpt-4o-with & 1.662 & 1.445 & 1.811 & 1.511 & 1.763 & 1.475 \\
gpt-4o-no & 1.488\textbf{*} & \textbf{1.369*} & 1.339\textbf{*} & \textbf{1.293*} & \textbf{1.156*} & 1.187 \\
\midrule
\multicolumn{7}{c}{\textbf{Baseline Rating Results}} \\
\midrule
TGT & 3.941 & 4.221 & 4.323 & 4.445 & 4.102 & 4.289 \\
CMF & 1.598 & 2.061 & 1.218 & 1.567 & 1.627 & 2.136 \\
EMCDR & 1.557 & 1.992 & 1.422 & 1.685 & 2.299 & 2.735 \\
PTUPCDR & \textbf{1.182} & 1.571 & \textbf{1.016} & 1.312 & 1.711 & 2.376 \\
\bottomrule
\end{tabular}
}
\setlength{\belowcaptionskip}{-15pt} 
\caption{LLM and baseline results for rating prediction task across CDR pairs. \textbf{Bold} indicates the best performance and \textbf{*} indicates best LLM performance.}
\label{tab:llm_rating_results}
\end{table*}
}

\subsection{RQ3: Medium versus High Context Prompt}
\label{sec:medium_high_context_results}

In this section, we examine the impact of additional context in prompting LLMs. We used the best-performing LLM from our results—either GPT-4 or GPT-4o, depending on the pair—and compared medium context prompts with high context prompts. The results are shown in Figures \ref{fig:ranking_context_results} and \ref{fig:rating_context_results}. The medium context prompts provided minimal information about cross-domain recommendation, lacking clarity on what to expect and had no background on the task. In contrast, the high context prompts included detailed background information about cross-domain recommendation, explicitly stated the available domains, and clearly explained the expected output and reasoning process. This experiment was conducted for both the ranking and rating tasks to assess how different levels of contextual information affect LLM performance. To create the medium context prompts, we removed the initial paragraph that explained CDR, and hid the names of the domains from the original prompts shown in Figures \ref{box:prompt-with_injection_rating} and \ref{box:prompt-with_injection_ranking}.

The results indicate that the high context prompt outperformed the medium context prompt across all CDR domain pairs for both ranking and rating, except for Pair 3 in the ranking task. As discussed in Section \ref{sec:ranking_results_section}, the ranking results for Pair 3 were not better than the baselines, which is likely due to the dissimilar domain combination. This poor performance in Pair 3 may explain why the high context prompt performed worse than the medium context prompt, as the additional information could have introduced noise, potentially confusing the LLM. However, we show that LLMs are able to perform exceptionally well when more context is added to the prompt, highlighting the fact that additional knowledge and guidance can lead to better results in CDR. 
{\scriptsize
\begin{figure*}[t]
    \centering

    \begin{subfigure}{.33\textwidth} 
        \centering
        \includegraphics[height=2cm]{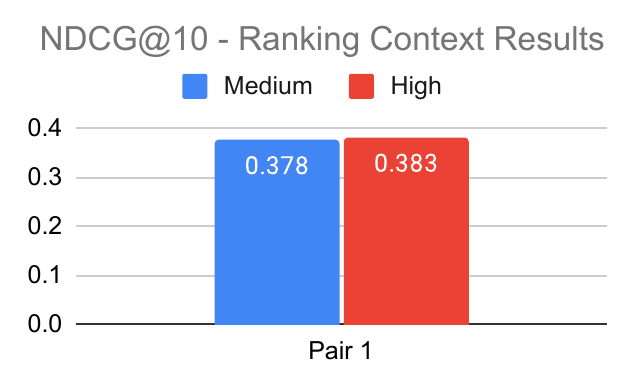}
    \end{subfigure}%
    \hfill
    \begin{subfigure}{.33\textwidth} 
        \centering
        \includegraphics[height=2cm]{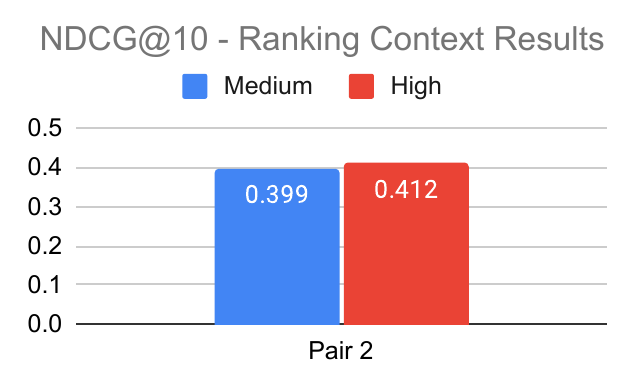}
    \end{subfigure}%
    \hfill
    \begin{subfigure}{.33\textwidth} 
        \centering
        \includegraphics[height=2cm]{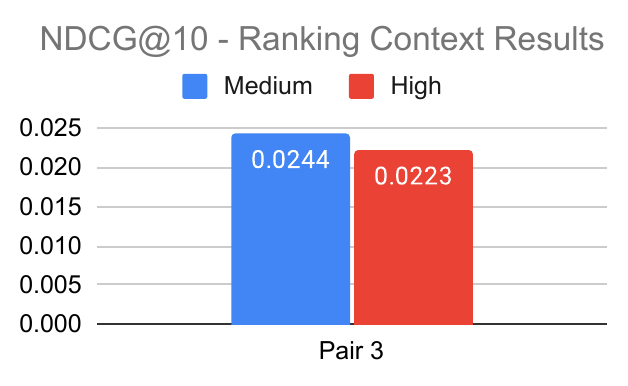}
    \end{subfigure}  
    \setlength{\abovecaptionskip}{-10pt}
    \setlength{\belowcaptionskip}{-12pt}
     \caption{LLM Ranking Results: Medium vs. High Context Prompts}
    \label{fig:ranking_context_results}
\end{figure*}
}
{\scriptsize
\begin{figure*}[t]
    \centering

    \begin{subfigure}{.33\textwidth} 
        \centering
        \includegraphics[height=2cm]{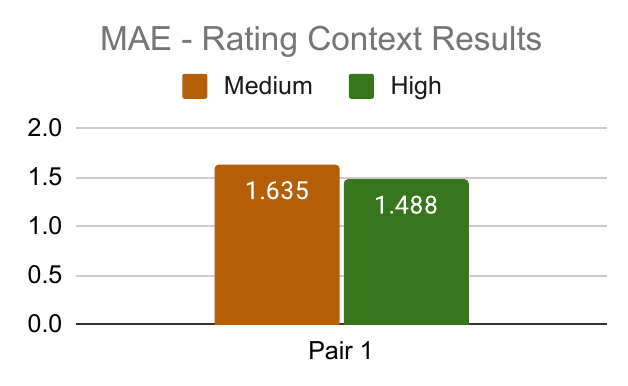}
    \end{subfigure}%
    \hfill
    \begin{subfigure}{.33\textwidth} 
        \centering
        \includegraphics[height=2cm]{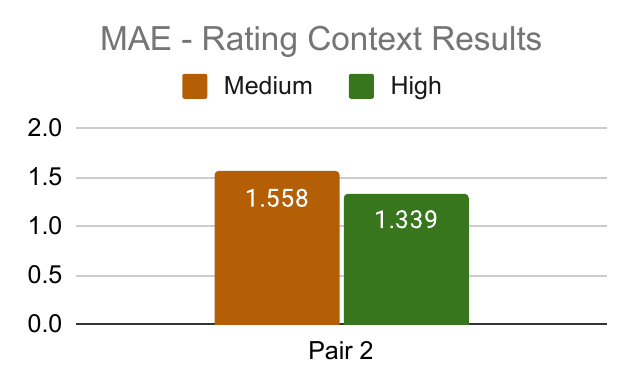}
    \end{subfigure}%
    \hfill
    \begin{subfigure}{.33\textwidth} 
        \centering
        \includegraphics[height=2cm]{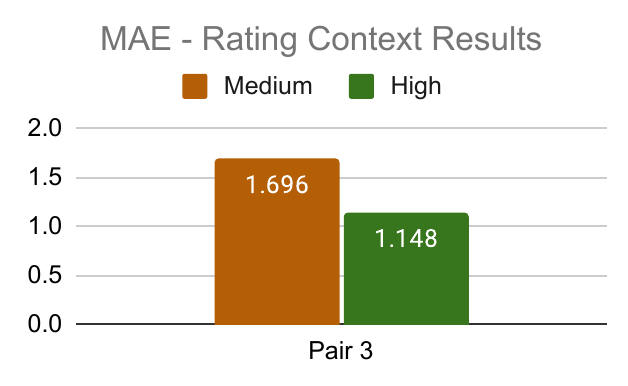}
    \end{subfigure}
    \setlength{\abovecaptionskip}{-10pt}
    \setlength{\belowcaptionskip}{-15pt}
    \caption{LLM Rating Results: Medium vs. High Context Prompts}
    \label{fig:rating_context_results}
\end{figure*}
}
\section{Related Work}

CDR has emerged recently as a promising solution to alleviate the cold-start and sparsity issues that single-domain recommender systems face \cite{zhu2021cross,vajjala2024analyzing,zang2022survey}. Recent models such as EMCDR \cite{man2017cross} and PTUPCDR \cite{zhu2022personalized} introduced mapping functions, where user embeddings from the source domain are transformed to work well in the target domain. Neural network-based approaches like UniCDR \cite{cao2023towards} and DisenCDR \cite{cao2022disencdr} further extend this idea by incorporating shared and disentangled latent representations to model both domain-specific and cross-domain user preferences. While these methods achieve state-of-the-art results, they are often very complex, tailored to specific domain pairs, and rely on large amounts of interaction data, making them less effective in situations where data is sparse. This highlights the need for more adaptable and efficient solutions in CDR.

Recently, in the field of single-domain RS, there has been research that leverages the abilities of LLMs to make recommendations \cite{zhao2023recommender,wu2024survey}. For example, Chat-REC \cite{gao2023chat} prompts LLMs to improve explainability for recommendations in a single domain, showing promising performance compared to traditional RS methods. LLMRec \cite{lyu2023llm} has been introduced, where they used LLMs to augment the interaction graph for users and items, showing promising results. However, there has been very limited work in CDR that has leveraged LLMs to improve recommendation performance. Petruzzelli et al. \cite{petruzzelli2024instructing} introduced work that prompts LLMs for explainability in the CDR domain, but it focuses mainly on explainability and does not explore prompts tailored to specific CDR tasks, such as ranking or rating prediction. In contrast, our study provides an extensive evaluation of LLMs across multiple tasks, demonstrating their potential as competitive alternatives to CDR methods in both similar and dissimilar domain combinations.

\section{Conclusion and Future Work}

In this paper, we introduce novel prompts specifically designed for the task of CDR and demonstrate the capabilities of LLMs in performing CDR tasks. Through extensive evaluation across three CDR scenarios, we show that LLMs can outperform recent baseline methods in both ranking and rating prediction tasks across various domain combinations, which highlights their potential for the CDR domain. This work is significant as it showcases the ability of LLMs to enhance recommendation performance by effectively understanding and transferring user preferences across domains in CDR. For future work, we aim to look into developing hybrid models that combine the strengths of LLMs and traditional CDR methods, and we plan to explore dynamic prompts that adapt based on the specific domain combinations to optimize recommendations for CDR.

%
%
\bibliographystyle{splncs04}
\bibliography{bibliography}

\end{document}